# Acoustophoresis-based system for a high throughput cell washing: impact on mesenchymal stromal cells transcriptome

Hugo R. Sugier[1,2], Ludovic Bellebon[3], Jean-Luc Aider[3], Jérôme Larghero[4], Juliette Peltzer[5] and, Christophe Martinaud[6].

[1]Aenitis technologies, Paris, France
E-mail: hugo.sugier@aenitis.fr
[2]Institut André Lwoff, INSERM UMR-MD 1197, Villejuif, France
[3]Laboratoire PMMH, UMR7636 CNRS, ESPCI Paris - PSL, Paris Sciences Lettres, Sorbonne Université, Paris, France
[4]Université de Paris, Assistance Publique-Hôpitaux de Paris, Hôpital Saint-Louis, Paris, France; Unité de Thérapie Cellulaire, INSERM U976, Centre d'investigation clinique de Biothérapies CBT501, Paris, France
[5]Institut de Recherche Biomédicale des Armées, Clamart, France
[6]Centre de Transfusion Sanguine des Armées, Clamart, France

## Introduction

With the emergence of a wide variety of innovative biotherapies over the past decades, production of cells on a large-scale has become a major challenge. Among different therapies relying on cell manufacturing, one can point out monoclonal antibodies[1], recombinant proteins[2], extracellular vesicles[3,4], CAR-T cells[5], induced pluripotent stem cells (iPSCs)[6] and, mesenchymal stromal cells (MSCs)[7]. One of the greatest remaining challenges towards large-scale production is processes automation in order to reduce costs and to increase the accessibility of innovative biotherapies to patients[8]. Cell washing/medium exchange is a key practice in cell manufacturing and is conventionally performed by centrifugation. However, despite being widely used and documented, centrifugation is time-consuming, labor-intensive and, induces important physical stress on cells[9,10]. In addition, centrifugation is a difficult method to integrate in a full automated process and presents a high risk of contamination, hence, making it a non-optimized approach for large-scale cell production.

To overcome limitations of centrifuged-based technologies, microfluidics has emerged as a promising solution for handling cells in a continuous and controlled flow. Several passive approaches have been developed such as deterministic lateral displacement[11–13], hydrodynamic filtration[14–16], pinched flow fractionation[17,18] and, inertial microfluidics[19–21]. However, the main limitation of passive means remains flow rate and throughput, mainly due to the flow conditions and microchannel shapes. To surmount those limitations, active methods can be applied based on different physical forces such as magnetic[22,23], optical[24], dielectrophoretic[25–27] and, acoustic[28].

Acoustic forces applied in a fluidic system (i.e., acoustophoresis) allow the manipulation of particles and cells. Acoustophoresis is label-free, biocompatible, induces low forces and physical stress and, can be easily integrated in an automated process, making it a promising and good manufacturing practices (GMP) compliant technology. Recently, acoustophoresis has been used for cell focusing[29], cell separation[30], transfection[31], measurement of cell physical properties[32] and, cell washing[33].

In 2004, Hawkes *et al.* have developed the first acoustic-based cell washing device with 2 inlets and 2 outlets. The authors were able to separate yeasts from a sodium fluorescein solution with a transfer of yeast up to 60% and a removal of fluorescein about 85% with a throughput of 1.7 mL/min [34]. Since this first proof of concept for medium exchange based on acoustophoresis, various other devices have been developed. Peterson et al., performed washing of bovine red blood cells (RBCs) with a throughput of 0.1 mL/min. RBCs recovery was up to 95% while 98% of the plasma contaminants were removed[35]. With a channel displaying a pre-focusing and a focusing zone, Tenje et al., were able to recover 97% of RBCs with no remaining albumin or IgA measurable[36]. Moreover, by adjusting channel design and fine tuning of the acoustic power, it is possible to accomplish more complex manipulations. Indeed,

Destgeer et al. have developed a channel with 3 different streams. They were able to perform a multimedium exchange with particles[37]. With a regard towards bioprocessing, acoustophoresis cell washing has been applied with Jurkat cells suspended in DMSO to mimic thawing of frozen cells which require a rapid removal of DMSO to minimize its toxicity on cells[38]. The cell recovery, up to 90-100% at a flow rate between 0.85 to 1.98 mL/min, combine with a high cell viability demonstrated the relevancy and efficiency of this method. Furthermore, the authors discussed the scalability of their device to reach the clinical needs. Another example of application where a rapid change of medium is essential is cell transfection. Hsi et al. transferred human primary T cells from PBS to a transfection solution at a flow rate of 60µL/min and performed a medium exchange efficiency of 86% for a transfection efficiency up to 60% while cell viability was slightly impacted (5% compared to control)[39]. Recently, Adler et al. compared a classic cell radiolabeling protocol by centrifugation to the acoustophoresis cell washing device AcouWash[40]. Cell radiolabeling efficiency and removal of unbound components were similar to centrifugation-based protocol. Moreover, cell viability and IFNγ production function were maintained after processing. Despite strong proof of concept and a commercially available device, acoustophoresis cell washing does not yet meet the needs of cell manufacturing.

In this work, we present a system based on acoustic waves allowing the transfer of cells from one medium to another while limiting the transfer of proteins. We optimized flow rate in order to increase cell throughput. Then, we tested the limits of our system with different cell types and protein concentrations. Finally, we performed a cell washing of both RBCs and MSCs in a loop system and evaluated the impact of the processing on cell viability and MSCs' transcriptome.

## Acoustophoresis

Acoustophoresis is the application of an acoustic field in a fluidic system allowing the manipulation of particles or cells. Particles with a positive acoustic contrast factor ($\Phi$) exposed to an acoustic field will migrate towards the acoustic node following the acoustic radiation force, which is expressed as:

$$F_z^{rad} = 4\pi a^3 k E_{ac} \sin(2kz) \Phi(\rho, \kappa) \tag{1.1}$$

Where

$$k = \frac{2\pi}{\lambda_{ac}} \tag{1.2}$$

$$\Phi(\rho, \kappa) = \frac{1}{3}\left[\frac{5\rho_p - 2\rho_m}{2\rho_p + \rho_m} - \frac{\kappa_p}{\kappa_m}\right] \tag{1.3}$$

And a is the particle radius, k the wave number, $\lambda_{ac}$ the ultrasonic wavelength, $E_{ac}$ the acoustic energy density, z the distance from the acoustic node, $\rho_p$ and $\rho_m$ the density of the particle and medium respectively and, $\kappa_p$ and $\kappa_m$ the compressibility of the particle and medium respectively. As shown by Eq. (1.1), the acoustic radiation force depends on the particle size (radius cubed). As particles are suspended in liquid, $F_z^{rad}$ is balanced by Stokes drag force expressed as:

$$F_z^{drag} = -6\pi\eta a v_p \tag{1.4}$$

Where η is the dynamic viscosity and $v_p$ the velocity of the particle. The resulting velocity $v_p$ of the particle exposed to $F_z^{rad}$ can be expressed as:

$$v_p = \frac{2a^2 k E_{ac} \Phi}{3\eta} \sin(2kz) \tag{1.5}$$

The velocity of particles will be proportional to the square of their radius. Hence, cells (6 µm < diameter < 30 µm) will migrate rapidly towards the acoustic node compared to proteins (< 5 nm), allowing a quick transfer of cells from the acoustic antinode to the acoustic node while protein motion can be neglected.

## Materials and methods

### Acoustic setup

The experimental device is composed of a stainless-steel channel with 3 inlets and 3 outlets. The channel has a length of 130 mm long, a width of 23 mm and, an inner thickness of 250 µm. A piezo composite (50 x 10 mm) (Smart Material), controlled by an electrical generator AFG1022 (Tektronix), resonates at ~2.5MHz. Temperature of the system is regulated at +20 °C by a liquid cooling system composed of Peltier thermoelectric module DT-AR-075-24 (Adaptative thermal management) linked to a liquid loop system, a type K thermocouple (National Instruments), a programmable voltage supply (Tenma) and, a PID control loop system. The flow rates are controlled by 3 independent peristaltic pumps (Minipuls 3, Gilson) in silicon tubing (2.06 mm inner diameter, Gilson). A scheme of the acoustic setup is presented in **Fig 1**.

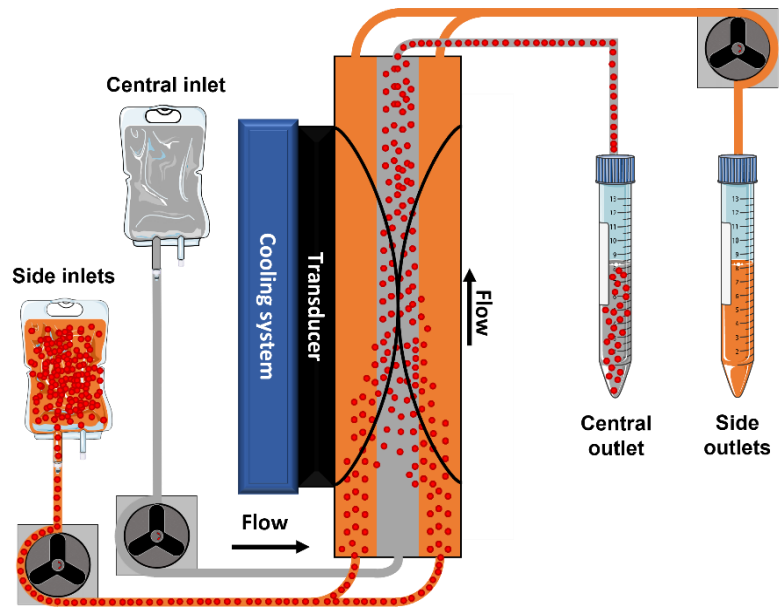

**Figure 1 Schematic of the acoustophoresis cell washing system.** Cells suspended in albumin solution were injected in the side inlets and physiological serum in the central inlet. When the cells pass in the acoustic field (represented with black waves) generated by the transducer, the cells undergo the acoustic radiation force and migrate towards the acoustic node at the center of the channel. Finally, the cells were collected into the central outlet, suspended in physiological serum.

### Stream mixture assessment

To assess the streams mixture, we injected MEM-α (Biological Industries), a culture media, in the side inlets and physiological serum (Fresenius Kabi) in the central outlet. The stream mixture was estimated by spectrometry measuring the absorbance of the central outlet with a PR 4100 Absorbance Microplate Reader (Bio-Rad). The percentage of MEM-α transferred to the central stream was estimated by its absorbance at 450 nm which follows a linear curve y = 442.78x – 14.405, $R^2$ = 0.9999.

### RBCs

RBC concentrates were obtained from healthy volunteer after written informed consent. According to French law, a declaration but no ethic comity approval was required for using these samples. Whole blood was collected in Vacutainer® EDTA K2E tubes (BD Biosciences), centrifuged and, RBCs were used within the following 4 h.

**Cell recovery**

Cell concentration was measured by flow cytometry using a CytoFLEX (Beckman Coulter) equipped of peristaltic pumps allowing the measure of a number of events (i.e., cells) per a defined volume. Data were analyzed with FCS Express 7 (De Novo Software). The percentage of cell recovery from the side inlets to the central outlet was estimated as following:

$$\% \ cell \ recovery = \frac{[\text{Central outlet}]_{cells} \ Q_C}{[\text{Central outlet}]_{cells} \ Q_C + [\text{Side outlets}]_{cells} \ Q_S} \times 100\% \quad (2)$$

Where $Q_C$ and $Q_S$ stand for central and side flow rates, respectively.

**Albumin removal**

Albumin concentration was measured using a BCG Albumin Assay Kit (Sigma-Aldrich) as recommended by the manufacturer. The percentage of albumin removal from the side inlets to the central outlet was estimated as following:

$$\% \ albumin \ removal = 100\% - \frac{[\text{Central outlet}]_{Albumin}}{[\text{Input}]_{Albumin}} \times 100\% \quad (3)$$

Prior measurement, the samples containing RBCs or MSCs were centrifuged for 5 min at 1200g and 450g, respectively, to avoid cells suspended interfering the absorbance.

**Separation efficiency**

The separation efficiency of cells and albumin was evaluated as following:

$$Separation \ efficiency = \frac{\% \ cell \ recovery}{\frac{[\text{Central outlet}]_{Albumin}}{[\text{Input}]_{Albumin}}} \quad (4)$$

**Apoptosis assay**

Cell viability was assessed by flow cytometry using a CytoFLEX (Beckman Coulter) with a FITC Annexin V (BD Pharmingen) staining, as recommended by the manufacturer.

**AD-MSCs**

Adipose tissue-derived MSCs (AD-MSCs) were collected from donors after written informed consent, undergoing a liposuction (Percy Military Medical Center, France). Adipose tissue was washed 3 times with DPBS (Corning) and enzymatically digested for 45 min at +37 °C under agitation with 0.075 mg/dL of collagenase type I (Sigma-Aldrich). Enzymatic digestion was stopped with 90% MEM-α and 10% human albumin (Vialebex 200 mg/mL). Cells were centrifuged and cell pellet was subsequently filtered at 100 μm and 30 μm. The resulting cells compose the stromal vascular fraction (SVF) which contains AD-MSCs. SVF was plated at 20 000 nucleated cells/cm² in MEM-α (Biological Industries) supplemented with 5% pooled human platelet lysate (French military blood center, France) and, 0.5% Ciprofloxacine (Bayer Pharma). The cells were incubated at +37 °C and 5% $CO_2$. After 24 hours, non-adherent cells were discarded to isolate adherent AD-MSCs. When 80% confluence was reached, cells were detached with trypsin-EDTA (TrypZean™ Solution, 1×, Sigma-Aldrich) for passage 1 (P1). After P1, cells were cryopreserved in MEM-α supplemented with 10% DMSO (Sigma-Aldrich) and 10% human albumin

(Vialebex 200mg/mL). After thawing, cells were plated at 4000 cells/cm² and harvested when confluence reached 80% at P2. MSCs surface antigen phenotype was confirmed by flow cytometry (Supplementary Table 1), following the minimum criteria defined by the International Society for Cellular Therapy[41]. The list of the antibodies used are provided in Supplementary Table 2. In all the following experiments, AD-MSCs were used at P3.

**Experimental procedure for cell washing with a loop**

RBCs and AD-MSCs were suspended at 1% hematocrit and $10.10^6$ cells/mL, respectively, in physiological serum supplemented with 5g/dL human albumin (Vialebex 200mg/mL) prior their injection into side inlets. RBCs and AD-MSCs were collected in the central outlet after their first passage into the loop (L1) for cell recovery, albumin removal and viability measurement. After their second passage into the loop (L2), the same investigations were conducted. In addition, AD-MSCs were collected after L2 for transcriptomic analysis. During all the experiments, cells were kept at +20 °C.

**Transcriptomic analysis**

**Statistical analysis**

All result measurements are expressed as mean ±SD resulting from 3, 4 or, 5 independent experiments. Normality of dataset was tested with Shapiro-Wilk test. Datasets which passed the normality test were challenged with ordinary one-way ANOVA or RM one way ANOVA tests while the others were challenged with Kruskal-Wallis or Friedman tests. All analyzes were performed with GraphPad Prism 9.3.0. and *P* values < 0.05 were considered significant.

## Results

**Flow rate configuration**

We first investigated which type of flow rate configuration would be the most suitable for a high cell recovery while avoiding a mixture of the different streams. We injected cell culture media (MEM-α) in the side inlets and physiological serum in the central inlet, two solutions with a close density (i.e., xxxx and xxxx, respectively). The lowest transfer of MEM-α was observed with a flow configuration $Q_S < Q_C$, with and without acoustics (**Fig 2a**). We then injected RBCs suspended in a 1% albumin solution in the side inlets and physiological serum in the central inlet. The highest albumin removal has been reached with $Q_S < Q_C$ (mean ±SD 31.72 ±SD 4.43), which was significantly different from $Q_S > Q_C$ (17.91 ± 4.86, *p*=0.0134) (**Fig 2b**). The RBC recovery was significantly more important with $Q_S < Q_C$ than $Q_S > Q_C$ or $Q_S = Q_C$ (95.97 ±0.65, 91.19 ±1.16 and, 93.82 ±0.62, *p*<0.0001 and *p*=0.0144 respectively) (**Fig 2c**). From Eq. (4), we calculated the separation efficiency of our system by compelling the cell recovery with the albumin removal resulting in a higher value for a more efficient cell washing (**Fig 2d**). The highest separation efficiency has been obtained with $Q_S < Q_C$ (1.41 ±0.09) and was significantly higher than $Q_S > Q_C$ (1.11 ±0.06, *p*=0.0071). Taken together, those results strongly suggest that the flow rate configuration the more adapted to our cell washing setup is $Q_S < Q_C$.

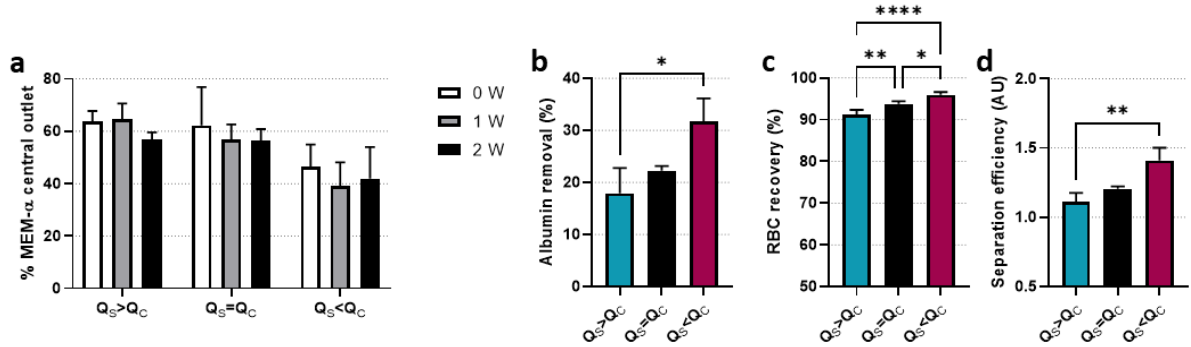

**Figure 2 Flow rate configuration $Q_S<Q_C$ is more suitable for cell washing.** (a) Percentage of MEM-α transferred from the side inlets to the central outlet depending on the flow rate configuration and acoustic power, n=3. (b) Percentage of albumin removal from a solution of RBCs suspended in 1g/dL of albumin depending on the flow rate configuration, n=4. (c) Percentage of RBC recovery from a solution of RBCs suspended in 1g/dL of albumin depending on flow rate configuration, n=4. (d) Separation efficiency of albumin and RBCs (arbitrary unit) from a solution of RBCs suspended in 1g/dL of albumin depending on flow rate configuration, n=4. All flow rates are specified in Supplementary Table 3. If not specified, power = 8 W, hematocrit = 0.02 and, albumin concentration = 1 g/dL. Histograms represent mean ±SD. *, ** and, ****stand for $p<0.05$, $p<0.01$ and, $p<0.0001$ with a parametric test, respectively.

**Optimization of flow rates**

Our first results led us to adopt a flow rate configuration where $Q_S < Q_C$. The next step was to optimize the flow rate in order to improve the separation efficiency. We first increased the central flow rate from 0.75 mL/min to 12 mL/min while keeping a side flow rate $Q_S$ = 0.75 mL/min. The highest albumin removal was obtained with $Q_C$ = 12 mL/min (89.31% ±0.53) which was significantly superior to $Q_C$ = 0.75 mL/min and $Q_C$ = 1.5 mL/min ($p<0.0001$) and, to $Q_C$ = 3 mL/min (81.41% ±0.59, $p=0.0419$) (**Fig 3a**). While the albumin removal was positively correlated with $Q_C$, the RBC recovery was less important at high flow rates. Indeed, the RBC recovery was about 87% at $Q_C$ = 0.75; 1.5; 3; 6 mL/min and decreased for $Q_C$= 9mL/min (83.07 ±0.97). The RBC recovery was significantly lower with $Q_C$ = 12 mL/min compared to all other flow rates (**Fig 3b**). Consequently, the separation efficiency was more important for $Q_C$ = 9 mL/min and $Q_C$ = 12 mL/min (6.85 ±0.24 and 6.90 ±0.11 respectively) (**Fig 3c**). Given the small difference in separation efficiency between $Q_C$ = 9 mL/min and $Q_C$ = 12 mL/min, we chose to work at $Q_C$ = 9mL/min, which allows a lower consummation of solution. In order to increase the throughput of cells, we increased the side flow rate from 0.75 mL/min to 9 mL/min. Both the albumin removal and the RBC recovery were negatively correlated to $Q_S$ resulting in a lower separation efficiency while $Q_S$ was increased (**Fig 3d**, **e** and, **f**). From those results, we decided to apply a flow rate configuration $Q_C$ = 9 mL/min and $Q_S$ = 0.75 mL/min for the following experiments.

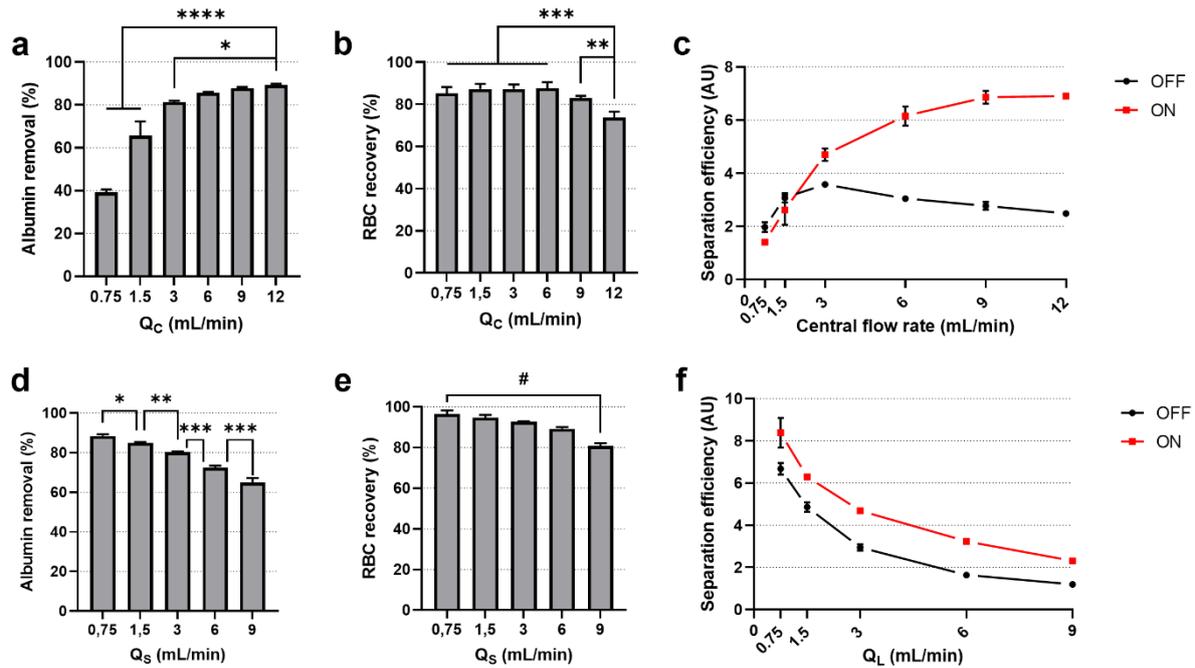

**Figure 3 Optimization of the flow rates for cell washing. (a)** Percentage of albumin removal from a solution of RBCs suspended in 1g/dL of albumin depending on $Q_C$. **(b)** Percentage of RBC recovery from a solution of RBCs suspended in 1g/dL of albumin depending on $Q_C$. **(c)** Separation efficiency calculated of albumin and RBCs from a solution of RBCs suspended in 1g/dL of albumin depending on $Q_C$. The black and red line show mean ±SD with the acoustics turned off or on, respectively. **(d)** Percentage of albumin removal from a solution of RBCs suspended in 1g/dL of albumin depending on $Q_S$. **(e)** Percentage of RBC recovery from a solution of RBCs suspended in 1g/dL of albumin depending on $Q_S$. **(f)** Separation efficiency of albumin and RBCs from a solution of RBCs suspended in 1g/dL of albumin depending on $Q_S$. The black and red line show mean ±SD with the acoustics turned 2off or on, respectively. If not specified, power = 8 W, hematocrit = 0.02, albumin concentration = 1 g/dL, $Q_S$ = 0.75 mL/min and, $Q_C$ = 9 mL/min. Histograms represent mean ±SD. *, **, *** and, **** stand for *p*<0.05, *p*<0.01, *p*<0.001 and, *p*<0.0001 with a parametric test, while # stands for *p*<0.05 with a nonparametric test, n=3.

**Impact of acoustic power, protein and, cell concentrations on separation efficiency**

We assessed the impact of different parameters on the cell washing performance: acoustic power, albumin concentration and, RBC hematocrit. The acoustic power applied in our system had no impact on the albumin removal (**Fig 4a**). The RBC recovery was positively correlated to the acoustic power starting to reach 90% from 4 W (**Fig 4b**). Therefore, the separation efficiency seemed to be more important for high acoustic power until it reaches a plateau, with the highest value at 8 W (9.41 ±0.27) (**Fig 4c**). The albumin removal and the RBC recovery were both positively correlated to the albumin concentration (**Fig 4d, e**). Consequently, the separation efficiency was significantly more important at higher albumin concentration, notably at 5 g/dL compared to 1 g/dL (25.47 ±1.97 and 11.63 ±0.54 respectively, *p*<0.0001) (**Fig 4f**). To test the limit of our cell washing system, we increased the cell concentration from 0.02 to 2 % hematocrit. We did not observe no change in albumin removal, RBC recovery or, separation efficiency. Those results suggest that the setup is not impacted by changes in cell hematocrit even at 2% of hematocrit (**Fig 4g**, **h** and, **i**).

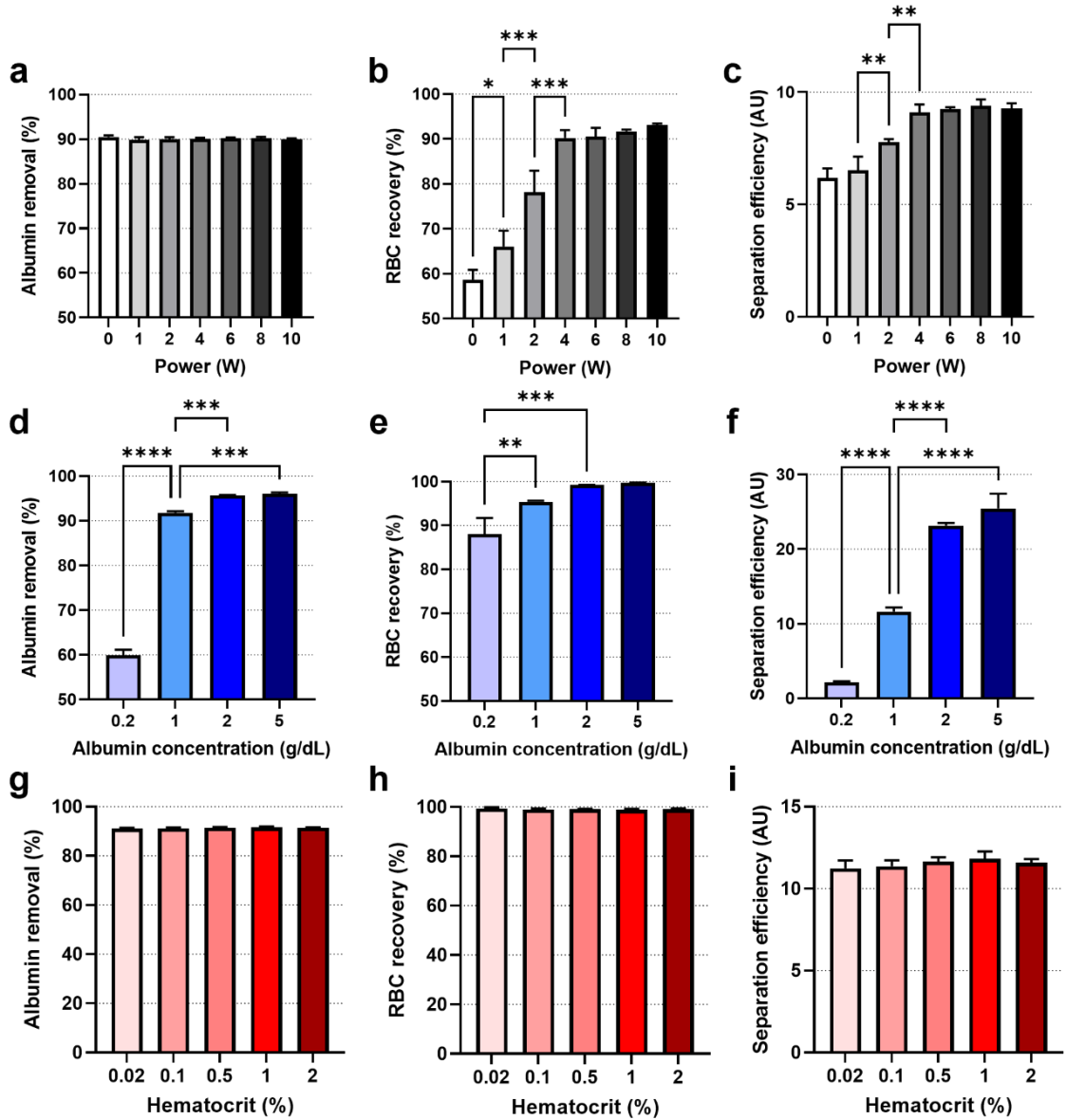

**Figure 4 Impact of acoustic power, albumin concentration and, hematocrit on cell washing.** (a) Percentage of albumin removal from a solution of RBCs suspended in 1g/dL of albumin depending on acoustic power. (b) Percentage of RBC recovery from a solution of RBCs suspended in 1g/dL of albumin depending on acoustic power. (c) Separation efficiency of albumin and RBCs from a solution of RBCs suspended in 1g/dL of albumin depending on acoustic power. (d) Percentage of albumin removal from a solution of RBCs suspended in 0.2 to 5 g/dL of albumin. (e) Percentage of RBC recovery from a solution of RBCs suspended in 0.2 to 5 g/dL of albumin. (f) Separation efficiency of albumin and from a solution of RBCs suspended in 0.2 to 5 g/dL of albumin (g) Percentage of albumin removal from a solution of RBCs suspended in 1g/dL of albumin depending on hematocrit. (h) Percentage of RBC recovery from a solution of RBCs suspended in 1g/dL of albumin depending on hematocrit. (i) Separation efficiency of albumin and RBCs from a solution of RBCs suspended in 1g/dL of albumin depending on hematocrit. All experiments were performed with $Q_C$ = 9 mL/min and $Q_S$ = 0.75 mL/min. If not specified, power = 8 W, hematocrit = 0.02, and albumin concentration = 1 g/dL. Histograms represent mean ±SD. *, **, *** and, **** stand for *p*<0.05, *p*<0.01, *p*<0.001 and, *p*<0.0001 respectively, n=3.

**Loop washing of RBCs and AD-MSCs**

To increase the proportion of protein removal, we processed the cells twice in a row (loop washing) in our cell washing facility and analyzed its performance after the first (L1) and second (L2) passage through the system. After loop washing of RBCs suspended in 5g/dL of albumin (n=5), we were not able to measure the albumin concentration of 3 samples (detection limit = 0.05 g/dL) while the 2 others were at a concentration of 0.08 g/dL (**Fig 5a**). The RBC recovery after L1 and L2 were both nearly at 100% (**Fig 5b**). Then, we assessed the proportion of apoptotic cells with an annexin staining. After L1, no difference was detected between the input and the cells processed with or without acoustics. We observed a significant increase of the proportion of apoptotic cells between L1 0W and L2 0W ($p$=0.0122) and, between L1 8W and L2 8W($p$=0.0138). On the other hand, no difference was detected between L1 0W and L1 8W or, between L2 0W and L2 8W (**Fig 5c**). After loop washing of AD-MSCs suspended in 5d/dL of albumin (n=3), we were not able to measure the albumin concentration in our samples processed at 8 W meaning that the remaining concentration was inferior to 0.05 g/dL (**Fig 5d**). The AD-MSCs recovery after L1 was about 99.5% after L2 at 8 W (**Fig 5e**). Finally, the ratio of apoptotic cells tended to be superior after L2 with or without acoustic compared to the input. It seemed that there was no difference of viability between the cells exposed to acoustics (8 W) or not ($p$=0.9887, n=3*)* (**Fig 5f**). Those results show that a second passage through the washing system allow a very high protein removal (≥99%) while allowing a high cell recovery (≥99%). Moreover, the shear stress induced a slight increase of apoptosis, but not the acoustics exposure.

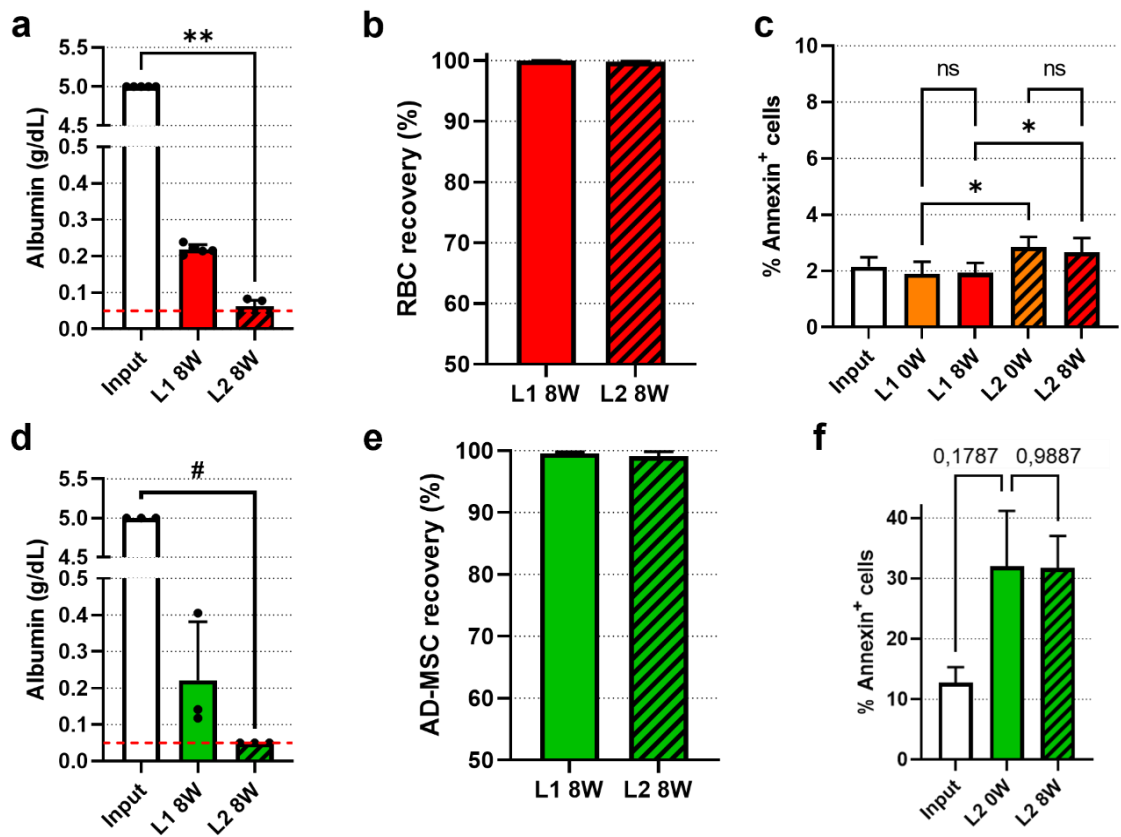

**Figure 5 Efficiency and impact on cell viability of loop washing of RBCs and AD-MSCs. (a)** Albumin concentration in the central outlet after loop washing from a solution of RBCs suspended in 5g/dL of albumin, n=5. The red dotted line stands for the detection limit of albumin. Values outside the detection range were plotted at 0.05 g/dL. **(b)** Percentage of cell recovery after loop washing from a solution of RBCs suspended in 5g/dL of albumin, n=5. **(c)** Percentage of annexin positive cells after loop washing from a solution of RBCs suspended in 5g/dL of albumin, n=5. **(d)** Albumin concentration after loop washing of AD-MSCs suspended in 5g/dL of albumin, n=3. The red dotted line stands for the detection limit of albumin. Values outside the detection range were plotted at 0.05 g/dL. **(e)** Percentage of cell recovery after loop washing of AD-MSCs suspended in 5g/dL of albumin, n=3. **(f)** Percentage of annexin positive cells after loop washing of AD-MSCs suspended in 5g/dL of albumin, n=3. All experiments were performed at $Q_C$ = 9 mL/min and $Q_S$ = 0.75 mL/min. RBCs and AD-MSCs

were suspended at 1 % hematocrit and 10 .10$^6$ cells/mL respectively. Histograms represent mean ±SD. * and ** stand for *p*<0.05 and *p*<0.01 respectively with a parametric test, while # stands for *p*<0.05 with a nonparametric test.

**Impact of loop washing on MSC transcriptome**

Prévision de la réception des analyses de résultats de RNA-seq : 26/09/2022

## Discussion

Requirement for cell production can be very different depending on the application. For allogeneic products (i.e., primary cells collected from one patient for a product destined to different patients), the maximum processing capacity is the key feature to take in consideration because of volume ranges which can vary from 50 mL[42] to 20,000 L[43]. Large volume treated expects a high throughput in order to avoid any negative impact on cells due to change of temperature and pH or, necessity to quickly remove toxic agents (e.g., cryopreservative and transfection agents). Moreover, it is essential to limit processing time between the first cells handled and the last to avoid cell variability into the batch. In contrast, autologous cell production is more limited in terms of volume, especially with pediatric application. In this case, volume and number of cells are very small. The associated challenges will be to ensure a high recovery rate with low minimum operating volume (i.e., avoid dead volumes due to tubing). Despite a variety of wash system devices, reviewed recently by Li et al.[44], there is still a lack of flexible technology that can handle small samples and provide a throughput consistent with large-scale needs.

The aim of this study was to develop a biocompatible medium exchange system allowing a high throughput which could be easily integrated in an automated and enclosed process. Based on acoustic waves, our system allows cell washing at a flow rate of 0.75 mL/mn in a single channel. Cell recovery was around 100% while removing up to 90% of surrounding proteins. In order to increase the cell washing efficiency, we performed a loop washing. With a second passage through the acoustic channel, we removed up to 99% of protein while cell recovery was about 99%. We tested the limit of our system by increasing RBC concentration up to 2% of hematocrit. We did not observe any decline in separation efficiency. Furthermore, the higher the protein concentration, the better our separation efficiency.

Acoustophoresis is a promising solution which can be easily integrated to a continuous process. 4 different acoustic-based devices have already been developed for cell washing: Biosep acoustic cell retention system, Cadence acoustic separator, ekko cell processing system and, AcouWash. Each of them present advantages towards specific applications[40,44]. Our goal was to present an easy-to-use setup which can be flexible and scalable depending on the application.

Finally, the quality of processed cells remains a top priority for clinical application. We evaluated the impact of our process on cell viability of RBCs and MSCs. RBCs viability was not affected by a single passage through the device but only after a second passage with a low percent of apoptotic cells (less than 3%). The acoustic exposure did not induce any change of viability. MSCs, which are non-circulating cells were more impacted by the shear stress. Indeed, after 2 passages through the channel, the percent of apoptotic cells went from 10% (input) to 30% (output). As with RBCs, the acoustic exposure did not increase the percent of apoptotic cells. These results are not surprising and could be explained by shear stress induces by the difference in flow rate between lateral and central inlets. While RBCs are physiologically used to flow rates about 3 to 26 mL/mn in arteries[45], AD-MSCs are resident cells which could explain the lower resistance to shear stress. Despite numerous works demonstrating the biocompability of acoustophoresis[46–49], to our knowledge, our study is the first exploring the impact of acoustophoretic manipulation on the whole transcriptome.